\documentclass[multphys,vecphys]{svmult}

\usepackage{makeidx}         % allows index generation
\usepackage{graphicx}        % standard LaTeX graphics tool
\usepackage{multicol}        % used for the two-column index
\usepackage[bottom]{footmisc}% places footnotes at page bottom
\makeindex             % used for the subject index
\begin{document}

\title*{Precision laboratory UV and IR wavelengths for cosmological and astrophysical applications}
\titlerunning{Precision laboratory wavelengths}

\author{M. Aldenius\and
S. Johansson}

\institute{Atomic Astrophysics, Lund Observatory, Box 43, SE-221 00 Lund, Sweden
\texttt{maria@astro.lu.se}}

\maketitle

\section{Introduction}\label{aldenius:sec1}
The quality of astronomical spectra is now so high that the accuracy of the laboratory data is getting more and more important. Both in astrophysics and in cosmology the needs for accurate laboratory wavelengths have increased with the development of new ground-based and air-borne telescopes and spectrographs. The high-resolution UV Fourier Transform (FT) spectrometer at Lund Observatory is being used for studying laboratory spectra of astrophysically important elements.

The ongoing investigations of possible variations in fundamental constants are demanding very accurate laboratory wavelengths of better than $\delta \lambda$\,$\sim$\,0.2\,m\AA . One of the methods of investigating such variations in the fine-structure constant, $\alpha\equiv (1/\hbar c)(e^{2}/4\pi \epsilon_{0})$, is the many-multiplet (MM) method, see e.g. \cite{aldenius1}. This method requires very accurate relative laboratory wavelengths for a number of UV resonance lines from several ionic species.

With the new focus on IR spectra within astrophysics (e.g. CRIRES at VLT)\index{CRIRES} the demands of accurate laboratory IR wavelengths have increased. Spectra from low density astrophysical plasmas (e.g. nebulae) contain parity-forbidden lines, which are of interest for diagnostics. These lines cannot in general be studied in laboratory spectra, but by using allowed transitions the values of the energy levels can be improved and new accurate wavelengths can be determined.
\begin{figure}
\centering
\includegraphics[angle=270, width=9.0cm]{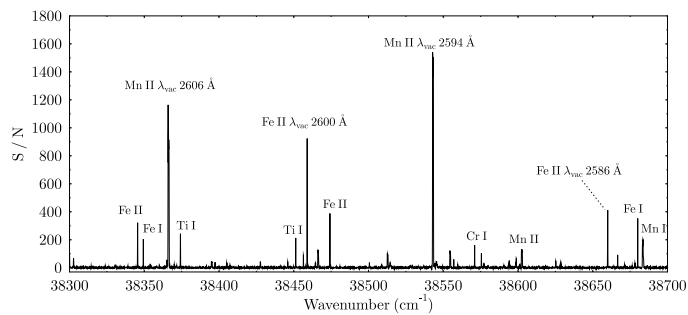}
\caption{A 400\,cm$^{-1}$ region of the spectrum, containing four of the UV lines}\label{aldenius:fig1}
\end{figure}

\section{Resonance UV wavelengths for cosmology}\label{aldenius:sec2}
Laboratory wavelengths and wavenumbers of in total 23 UV lines visible in high-redshift quasar absorption spectra have been measured using high-resolution FT spectrometry. To improve the accuracy of the relative wavelengths lines from Mg{\sc \,i}, Mg{\sc \,ii}, Ti{\sc \,ii}, Cr{\sc \,ii}, Mn{\sc \,ii}, Fe{\sc \,ii} and Zn{\sc \,ii} have been measured simultaneously, using a composite light source, see Fig.~\ref{aldenius:fig1}. Emphasis has been put on wavelength calibration and possible line structure, such as isotope structure and hyperfine structure. The uncertainties of the absolute and relative wavelengths are estimated to be 0.1-0.2\,m\AA\ and 0.03\,m\AA , respectively. The results have been published in \cite{aldenius2}.

\section{Parity forbidden IR wavelengths for astrophysics}\label{aldenius:sec3}
\begin{figure}
\centering
\includegraphics[width=5.9cm]{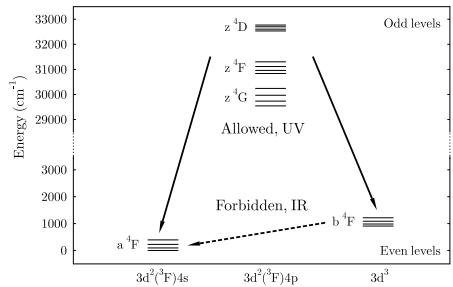}
\caption{Partial energy level diagram of Ti{\sc \,ii} displaying low even levels and higher odd levels, with parity forbidden IR lines and allowed UV lines}\label{aldenius:fig2}
\end{figure}

In low density astrophysical plasmas (e.g. nebulae) low lying metastable states can be populated. The possible radiative decay channels for these are through parity forbidden, M1 and E2, infrared transitions. Since these lines in general not can be observed in laboratory spectra the measured UV spectra, complemented with more measurements, are used for determining accurate values of the energy levels involved. As many UV transitions as possible are used to improve the accuracy of the energy levels, see Fig.~\ref{aldenius:fig2}. The wavelengths for the forbidden lines are then determined using Ritz combination principle. Our measurements include IR lines from Ti{\sc\,ii}, Cr{\sc\,ii}, Mn{\sc\,ii} and Fe{\sc\,ii}.

\printindex
\end{document}